\documentclass[%
 reprint,
superscriptaddress,
showpacs,preprintnumbers,showkeys,
 amsmath,amssymb,
 aps,
pra,
]{revtex4-1}

\usepackage{mathtools}
\usepackage{graphicx}% Include figure files
\usepackage{dcolumn}% Align table columns on decimal point
\usepackage{bm}% bold math
\usepackage{ bbold }% \mathbb{}
\begin{document}

%\preprint{APS/123-QED}

\title{Quasiparticle spectra of supersolid lattice gases at near-resonant Rydberg-dressing}% Force line breaks with \\
%\thanks{A footnote to the article title}%

\author{Andreas Gei{\ss}ler}
\email{andreas.geissler87@gmail.com}
\affiliation{Institut f\"ur Theoretische Physik, Goethe-Universit\"at, 60438 Frankfurt/Main, Germany}
\author{Ulf Bissbort}
\affiliation{Singapore University of Technology and Design, 1286082 Singapore}%
\affiliation{Department of Nuclear Science and Engineering and Research Laboratory of Electronics, Massachusetts Institute of Technology, Cambridge, MA, USA}%
\author{Walter Hofstetter}
\affiliation{Institut f\"ur Theoretische Physik, Goethe-Universit\"at, 60438 Frankfurt/Main, Germany}

\date{\today}% It is always \today, today,
             %  but any date may be explicitly specified

\begin{abstract}

One of the major challenges in realizing a strongly interacting lattice gas using Rydberg states is the occurrence of avalanche loss processes. As these are directly proportional to the total Rydberg fraction, the commonly suggested solution is using far off-resonantly excited Rydberg states. We instead propose the realization of a correlated bosonic lattice gas at near-resonant excitation, where the total Rydberg fraction in the bulk is low due to the strong, interaction-driven effective detuning. Using real-space dynamical mean-field theory we show that its reduced effect at the boundary of a system can easily be compensated by considering a tailored beam-waist of the driving Rabi-laser. In this geometry we discuss the spectral properties at the crossover between the supersolid and the superfluid state and present the momentum resolved spectral properties of the supersolid bulk. The latter results are obtained within an extended quasiparticle method which also yields a correction of the mean-field phase transition.
 
\end{abstract}

% PACS
%\keywords{Rydberg dressing, bosonic, supersolid}%Use showkeys class option if keyword
                              %display desired
                              
\maketitle

%\tableofcontents

\paragraph*{Introduction}

Experiments in recent years have shown the feasibility of using Rydberg excitations to introduce long-range interactions to many-body quantum gas experiments, as a new element of the ultracold atom platform for quantum simulation of strongly correlated systems \cite{Hofstetter2018}. Such experiments have already verified the emergence of a dressed interaction potential \cite{Jau2015,Zeiher2016} as well as non-trivial collapse and revival dynamics \cite{Zeiher2017,Bernien2017} for far detuned driving of the excitation. In this regime there also exists a vast body of theoretical work on crystalline \cite{Lauer2012a,Hoening2014,Petrosyan2015a} and supersolid phases \cite{Pupillo2010c,Henkel2010c,Cinti2014}, some also discussing spectral properties \cite{Saccani2012,Ancilotto2013a,Macri2014}.

In contrast, the case of near-resonant Rydberg excitation has been studied to a far lesser extent \cite{Weimer2010,Pohl2010,Rademaker2013,Vermersch2015}, with only few works considering the itinerant dynamics needed for supersolid formation \cite{Saha2014,Geissler2017}. Also, some experiments have investigated the frozen case, finding signatures of strong correlations \cite{Gunter2013,Helmrich2018} or reconstructing the density matrix at low particle density \cite{Gavryusev2016a}. But most importantly, experiments in this regime have revealed a major obstacle for achieving itinerancy for coherently driven near-resonant Rydberg excitations, namely the blackbody radiation-induced avalanche loss \cite{Goldschmidt2016,Aman2016,Boulier2017}. On the timescale of itinerancy, it behaves as an instantaneous global process. Therefore, the relevant timescale $\tau$ is given by the onset of the avalanche due to blackbody radiation-induced transfers from an excited Rydberg level to any nearby level of opposite parity. As this can be triggered anywhere in the system, $\tau^{-1}$ is proportional to the total number of Rydberg excitations. Therefore, most dressing experiments preferentially focus on small and low-dimensional systems \cite{Schauss2015,Jau2015,Zeiher2017,Bernien2017}.

\begin{figure}[h]
 \centering
 %\captionsetup{width=0.95\textwidth}
 \includegraphics[width=0.98\columnwidth]{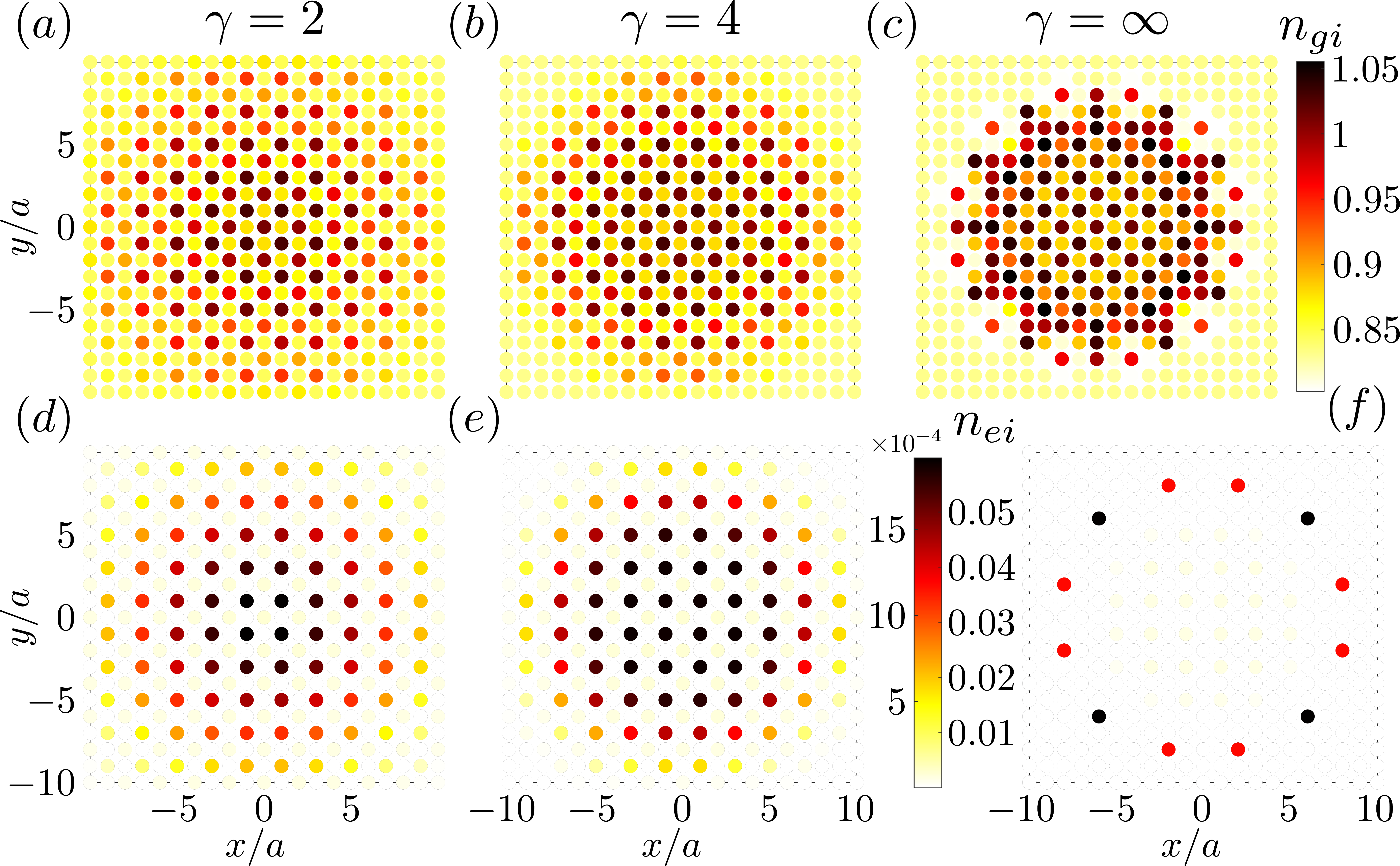}
 \caption{Supersolid formation with spatially confined Rydberg excitations. Real-space density distributions of ground and excited state components for various beam-waists $\gamma = 2 , 4 , \infty$ (see text). % in~\eqref{eq:beam-waist}.
  Note the different color axis for $(d),(e)$ and $(f)$. Using natural units ($\hbar = 1$), the parameters used in~\eqref{eq:full-hamil} are $ \lbrace U_{gg},U_{ge},U_{ee},J,\mu,\Delta,\Omega,V \rbrace = \lbrace 3\times 10^{-3},5,100,3\times 10^{-4},3\times 10^{-4},-0.2,0.2,10^4 \rbrace \left[ \textrm{MHz} \right] $.
 }
 \label{fig:trapped_SS}
\end{figure}

In this work we present an experimentally feasible method to significantly increase $\tau$ in a two-dimensional system and discuss the spectral properties of such a system at near-resonant Rydberg excitation. In a finite system with repulsive Rydberg interactions the bulk Rydberg fraction is vastly reduced due to the interaction-induced mean-field shift of the detuning \cite{Geissler2017}, while it becomes significant at the boundary. Thus, the avalanche is primarily triggered there. We therefore propose to exploit the spatial inhomogeneity of the driving laser to achieve a strong reduction of the Rydberg fraction in the boundary region, even below bulk values (see Fig.~\ref{fig:trapped_SS}).

%In this report we first review the Hamiltonian of the system, followed by an introduction of the method used to obtain the spectral properties in a generalized quasiparticle method. Then we discuss the formation of a supersolid using various Rabi beam-shapes with a focus on the obtained Rydberg fraction and corresponding spectral properties. In the conclusion we discuss the relevance of our results for current experimental setups.

\paragraph*{System}

To fully capture many-body effects resulting from near-resonantly driving a ground-to-Rydberg transition, we consider a two-component extended Hubbard model in the grand canonical ensemble, which  in terms of bosonic creation (annihilation) operators $\hat{b}^{\dag}_{\sigma i}$ $(\hat{b}_{\sigma i})$ has the form (for $\hbar = 1$)

\begin{equation} \label{eq:full-hamil}
\hat{H} = - J \sum_{\substack{\langle i,j \rangle , \sigma}} (\hat{b}^{\dag}_{\sigma i} \hat{b}_{\sigma j} + \textrm{h.c.} ) + \sum_{\substack{i,\sigma, \eta}} E_{\sigma\eta}^{i} \hat{b}^{\dag}_{\sigma i} \hat{b}_{\eta i} + \hat{H}_V,
%\hat{H} = - J \sum_{\substack{\langle i,j \rangle \\ \sigma}} (\hat{b}^{\dag}_{\sigma i} \hat{b}_{\sigma j} + \textrm{h.c.} ) + \sum_{\substack{\sigma, \eta \\ i}} E_{\sigma\eta}^{i} \hat{b}^{\dag}_{\sigma i} \hat{b}_{\eta i} + \hat{H}_V,
\end{equation}
with $\sigma,\eta = g,e$ for ground and Rydberg-excited state respectively, where we consider the tunneling with rate $J$ between all nearest neighbors $\langle i,j \rangle$ of a square lattice. Locally, we have the chemical potential $\mu = -E_{gg}^{i} $ and $E_{ee}^{i} = -\mu -\Delta$, where $\Delta$ is the effective detuning of the laser driving the Rydberg excitation with a Rabi frequency $ \Omega_i /2 = E_{ge}^{i} = E_{eg}^{i}$ that generally has a spatial dependence due to the beam-waist. With $\hat{n}_{\sigma i} = \hat{b}^{\dag}_{\sigma i} \hat{b}_{\sigma i}$ the interaction terms are 

\begin{equation}
\hat{H}_V = \sum_{\substack{i,\sigma, \eta }} \frac{U_{\sigma \eta}}{2} \hat{n}_{\sigma i} (\hat{n}_{\eta i} - \delta_{\sigma , \eta}) + V \sum_{i \neq j} \frac{\hat{n}_{ei} \hat{n}_{ej}}{d_{ij}^6}.
\end{equation}
The first terms include the various local inter- and intra-species Hubbard interactions. In the ground state these stem from short-range s-wave scattering processes, while we consider large values for $U_{ge} = U_{eg}$ and $U_{ee}$ (both $\gg U_{gg}$) due to the expected quantum Zeno blocking \cite{Vidanovic2014} of loss channels involving the local formation of Rydberg molecules \cite{Geissler2017} as strong dipole interactions dominate the short-range behavior. The Rydberg excited atoms exhibit a repulsion $V = C_6 / a^6$, given by the van der Waals constant of the respective Rydberg state (e.g. $C_6 = 241.6 \textrm{ MHz }\mu \textrm{m}^6$ for $36S$ of $^{87}$Rb \cite{Singer2005,Li2018}), and $a = 532 \, \textrm{nm}$ is the lattice spacing and length scale of the distance $d_{ij} = |\mathbf{r}_i - \mathbf{r}_j| / a$.

\paragraph*{Operator-based quasiparticle expansion}

To derive momentum-resolved quasiparticle (QP) properties we use an operator-based expansion of~\eqref{eq:full-hamil} which can be understood as an extension of Bogoliubov theory \cite{Bogolyubov1947} %(also discussed in \cite{Pitaevskii2003,Pethick2008}) 
beyond the limit of a single band of QP fluctuations \cite{Bissbort2012, Bissbort2014} also valid for strong interactions. It is based on linearized fluctuations of a variational  many-body ground-state wave function of Gutzwiller-type $| \psi_{\textrm{GW}} \rangle = \prod_i | \psi \rangle_i$ \cite{Rokhsar1991,Krauth1992}. In the following we introduce this method for states with partially broken lattice symmetries. %For an introduction of its original form assuming a homogeneous ground-state wave function we refer to \cite{Bissbort2012,Bissbort2014}.

We first define the fluctuation operators $\hat{\delta b}_{\sigma i} \equiv \hat{b}_{\sigma i} - \phi_{\sigma i}$ and $\hat{\delta n}_{\sigma i} \equiv \hat{n}_{\sigma i} - n_{\sigma i}$. % for the annihilation operator and the local occupation number respectively.
One obtains an exact representation of the original Hamiltonian,
%
%\begin{align}
$ \hat{H} = \sum_{i} \hat{H}_{\textrm{GW}}^{i}\left( \lbrace \phi_{\sigma i} \rbrace , \lbrace n_{\sigma i} \rbrace \right)+ \Lambda \left( \hat{\delta}^2 \right). $
%\end{align}
Its first term is given by a sum of local mean-field (MF) Hamiltonians self-consistently inter-coupled by the MF values $\phi_{\sigma i} = \langle \hat{b}_{\sigma i} \rangle$ and $n_{\sigma i} = \langle \hat{n}_{\sigma i} \rangle$ in the tunneling and long-range interaction terms respectively. The last term contains all expressions of higher order in $\hat{\delta b}_{\sigma i}$ and $\hat{\delta n}_{\sigma i}$ necessary to recover the original Hamiltonian. %In our case it only contains terms of second order in the fluctuation operators. 
The distribution of all MF ground-state values $\lbrace \phi_{\sigma i} \rbrace $ and $ \lbrace n_{\sigma i} \rbrace$ are obtained by their iterative calculation %of  $\phi_{\sigma i} = \langle \hat{b}_{\sigma i} \rangle_0$ and $n_{\sigma i} = \langle \hat{n}_{\sigma i} \rangle_0$, where $\langle \cdot \rangle_0$ specifies the evaluation of the expectation value 
in the lowest eigenstates $| 0 \rangle_i$ of the local Hamiltonians.

The eigenstates of each $\hat{H}_{\textrm{GW}}^{i}$ constitute a local eigenbasis with eigenenergies $E_{n}^{(i)}$. Numerically we only consider the $N$ lowest states amounting to an effective truncation of the bosonic Fock basis. %Within these eigenstates the Gutzwiller MF ground state is given as
As $| \psi_{\textrm{GW}} \rangle = \prod_{i} | 0 \rangle_{i}$, we define corresponding local Gutzwiller raising and lowering operators

\begin{equation}\label{eq:Gutzwiller_Ops}
%\begin{aligned}
|n \rangle_{i}  {}_{i}\langle 0| \equiv  \sigma_{i}^{(n)^{\dagger}}, \qquad 
%|0 \rangle_{i}  {}_{i}\langle 0| &=  \mathbb{1} -  \sum_{i>0} \sigma_{i}^{(n)^{\dagger}} \sigma_{i}^{(n)}, \\
|0 \rangle_{i}  {}_{i}\langle n| \equiv  \sigma_{i}^{(n)}.
% & |n \rangle_{i}  {}_{i}\langle m| &=  \sigma_{i}^{(n)^{\dagger}} \sigma_{i}^{(m)}. \\
%\end{aligned}
\end{equation}
Due to the completeness of these eigenbases for $N \rightarrow \infty$, we can exactly rewrite all terms appearing in $\Lambda \left( \hat{\delta}^2 \right)$ as sums of products of~\eqref{eq:Gutzwiller_Ops}. %To do so we introduce the notation  $ \pmb{\sigma}_{i} = \left(\sigma_{i}^{(1)} ,  \ldots , \sigma_{i}^{(N)} \right)^{\textrm{T}}$ and  $ \pmb{\sigma}^{\dag}_{i} = \left(\sigma_{i}^{(1)^{\dag}} ,  \ldots , \sigma_{i}^{(N)^{\dag}} \right)^{\textrm{T}}$ for the Gutzwiller operators. %, so the fluctuations are given as $\hat{\delta b}_{\sigma,i} \left( \pmb{\sigma}_{i} , \pmb{\sigma}_{i}^{\dag} \right)$ and $\hat{\delta n}_{\sigma,i} \left( \pmb{\sigma}_{i} , \pmb{\sigma}_{i}^{\dag} \right)$.
Thus one obtains an exact expansion $\Lambda \left( \hat{\delta}^2 \right) = \mathcal{H}^{(2)} +  \mathcal{H}^{(3)}  +  \mathcal{H}^{(4)} $ up to fourth order in the Gutzwiller operators, where no first-order term appears due to the self-consistency conditions defining $\hat{H}_{\textrm{GW}}^{i}$. The second-order term yields the full spectrum of non-interacting QP excitations, while higher-order terms generate interactions among them. %, resulting in their broadening and renormalization. 
A sufficiently low concentration of QP excitations implies that the higher-order terms can be neglected. %In the proximity [!!!] of the many-body ground-state we may neglect these terms for a sufficiently low concentration of QP excitations. 

In order to bring $\mathcal{H}^{(2)}$ into a diagonalizable form that allows QP properties to be extracted, we need to consider the commutation relations of~\eqref{eq:Gutzwiller_Ops}. As the system exhibits a spontaneously broken lattice translational symmetry, splitting the system into $L_c$ unit cells with $N_c$ sites each, we are particularly interested in the commutation relations of the partial Fourier transformed operators $\tilde{\sigma}^{(n)}_{\mathbf{k},s} = L_c^{-1/2} \sum_{l} e^{-i\mathbf{k} \cdot (\mathbf{r}_l+\mathbf{r}_s)} \hat{\sigma}^{(n)}_{l,s}$ and their hermitian adjoints $\tilde{\sigma}^{(n)^{\dag}}_{\mathbf{k},s}$, which can be used to represent the Hamiltonian. Here, the position $\mathbf{r}_i = \mathbf{r}_l + \mathbf{r}_s$ of each site $i \equiv (l,s)$ is given by the position $\mathbf{r}_l$ of the unit cell $l$ and the relative position $\mathbf{r}_s$ inside the unit cell. We label equivalent sites in all unit cells by the representative index $s$. A short derivation reveals that the commutation relations of the Gutzwiller operators are approximately bosonic, with the only non-vanishing relations
$\langle \left[ \tilde{\sigma}_{\mathbf{k}',s'}^{{(m)}} , \tilde{\sigma}_{\mathbf{k},s}^{{(n)}^{\dag}} \right] \rangle = \delta_{n,m} \delta_{\mathbf{k},\mathbf{k}'} \delta_{s,s'} - \delta_{s,s'} R / L_c$,
%\begin{equation}
%\left[ \tilde{\sigma}_{\mathbf{k}',s'}^{{(m)}} , \tilde{\sigma}_{\mathbf{k},s}^{{(n)}^{\dag}} \right] =& \delta_{n,m} \delta_{\mathbf{k},\mathbf{k}'} \delta_{s,s'} - \frac{\delta_{s,s'}}{L_c} R^{(n,m)}_{\mathbf{k},\mathbf{k}'}(s). \label{eq:GWfluc_commutk_red} 
%\left[ \tilde{\sigma}_{\mathbf{k}',s'}^{{(n)}^{\dag}} , \tilde{\sigma}_{\mathbf{k},s}^{{(m)}^{\dag}} \right] =& \left[ \tilde{\sigma}_{\mathbf{k}',s'}^{{(n)}} , \tilde{\sigma}_{\mathbf{k},s}^{{(m)}} \right] = 0. \label{eq:GWfluc_commutk_zero}
%\end{equation}
%Here we introduce the operator $R^{(n,m)}_{\mathbf{k},\mathbf{k}'}(s)$ whose expectation value $R = \langle R^{(n,m)}_{\mathbf{k},\mathbf{k}'}(s) \rangle$ describes the deviation from bosonic behavior 
where $R = \langle R^{(n,m)}_{\mathbf{k},\mathbf{k}'}(s) \rangle$ describes the deviation from bosonic behavior. Its precise form is discussed in more detail in Appendix \ref{sec:Commutator_deviation}. The essential approximation of the QP method is to take $R \rightarrow 0$ requiring sparsely occupied fluctuation modes. This can be quantified \textit{a posteriori} via the fraction $\epsilon_{s}$ of modes populating Gutzwiller excitations at the representative sites $s$. It is given by $ \epsilon_{s} = \langle \mathbb{1}_s - |0 \rangle_{s}  {}_{s}\langle 0| \rangle$. %$= L_c^{-1} \sum_{\mathbf{k},\gamma} \sum_{n>0} |v^{(\mathbf{k},\gamma)}_{n,s(i)}|^2 $, where $s(i)$ is the index of site $i$ within the corresponding unit cell. 
One can show that $ \delta_{n,m} \epsilon_{s} < R / L_c < (1 +  \delta_{n,m}) \epsilon_{s} $. The corresponding figure of merit
%\begin{align}
$ \sum_s \epsilon_s / N_c = \epsilon < 6\%$ for all discussed cases. % = \frac{1}{L} \sum_{\mathbf{k},\gamma} \sum_{s,n>0} |v^{(\mathbf{k},\gamma)}_{n,s}|^2$
%\end{align}
%For all discussed cases we find $\epsilon < 6 \%$.

Using the notation %$ \pmb{\sigma} = \left( \pmb{\sigma}_1 , \ldots  , \pmb{\sigma}_L \right)^{\textrm{T}}$ and $ \pmb{\sigma}^{\dag} = \left( \pmb{\sigma}^{\dag}_1 , \ldots  , \pmb{\sigma}^{\dag}_L \right)^{\textrm{T}}$ 
$ \pmb{\sigma} = \left( \sigma_{1}^{(1)} , \ldots  , \sigma_{L}^{(N)} \right)^{\textrm{T}}$ %and $ \pmb{\sigma}^{\dag} = \left( \sigma_{1}^{(1)^{\dag}} , \ldots  , \sigma_{L}^{(N)^{\dag}} \right)^{\textrm{T}}$
with corresponding Fourier-transformed vectors $\tilde{\pmb{\sigma}}$ %and $\tilde{\pmb{\sigma}}^{\dag}$
and for $R \rightarrow 0$ we find the approximate diagonal form

\begin{equation}
\mathcal{H}^{(2)} \approx {\sum_{\mathbf{k} \in 1.\textrm{BZ}',\gamma}} \omega_{\mathbf{k},\gamma} \beta_{\mathbf{k},\gamma}^{\dag} \beta_{\mathbf{k},\gamma} + \delta E_{\textrm{QP}}, \label{eq:HQP_diag_full}
\end{equation}
where the quasimomenta are confined to the reduced first Brillouin zone ($1.\textrm{BZ}'$) corresponding to the retained translational symmetry. We refer to Appendix \ref{sec:qp-Hamil-diag} for a detailed discussion of the diagonalization. The representation~\eqref{eq:HQP_diag_full} is given in terms of the generalized QP operators

\begin{equation}
\beta_{\mathbf{k},\gamma} \equiv \mathbf{x}^{(\mathbf{k},\gamma)^{\dag}} \Sigma \begin{pmatrix} \tilde{\boldsymbol{\sigma}} \\ \tilde{\boldsymbol{\sigma}}^{\dag} \end{pmatrix} \equiv \mathbf{u}^{(\mathbf{k},\gamma)^{\dag}}\tilde{\boldsymbol{\sigma}} + \mathbf{v}^{(\mathbf{k},\gamma)^{\dag}} \tilde{\boldsymbol{\sigma}}^{\dag}, \label{eq:QP_modeop1} 
%\beta_{\mathbf{k},\gamma}^{\dag} \equiv -\mathbf{y}^{(\mathbf{k},\gamma)^{\dag}} \Sigma \begin{pmatrix} \tilde{\boldsymbol{\sigma}} \\ \tilde{\boldsymbol{\sigma}}^{\dag} \end{pmatrix} \equiv \mathbf{v}^{(\mathbf{k},\gamma)^{T}}\tilde{\boldsymbol{\sigma}} + \mathbf{u}^{(\mathbf{k},\gamma)^{T}} \tilde{\boldsymbol{\sigma}}^{\dag}, \label{eq:QP_modeop2}
\end{equation}
obtained from the eigenvectors of the eigenvalue equations $\Sigma \tilde{\mathcal{H}}_{\textrm{QP}} \mathbf{x}^{(\mathbf{k},\gamma)} = \omega_{\mathbf{k},\gamma} \mathbf{x}^{(\mathbf{k},\gamma)}$, where $\Sigma = \textrm{diag} (\mathbb{1}_{NL} , - \mathbb{1}_{NL})$ and $L = N_c L_c$. By expressing~\eqref{eq:HQP_diag_full} with $\beta_{\mathbf{k},\gamma}^{\dag}$ and $ \beta_{\mathbf{k},\gamma}$ in normal order we find the scalar correction $\delta E_{\textrm{QP}}$ of the MF ground-state energy. It effectively lowers the total energy density in relation to its MF value $E_{\textrm{MF}} = \langle \psi_{\textrm{GW}} | \hat{H}| \psi_{\textrm{GW}} \rangle / L$, % due to the average shift of the mode energies in relation to the  Gutzwiller excitations, 
while it depends on the symmetry breaking of the MF ground-state. For nearly degenerate MF ground-states, close to a first-order phase transition, we thus obtain a correction to the location of the transition by comparing the energies $E_{\textrm{MF+QP}} = E_{\textrm{MF}} + \delta E_{\textrm{QP}} / L$.

To calculate dynamical correlation functions given in terms of linear combinations of local operators, such as the Fourier transforms $\tilde{b}_{\sigma \mathbf{k}}$ and $\tilde{n}_{\sigma \mathbf{k}}$ of $\hat{b}_{\sigma i}$ and $\hat{n}_{\sigma i}$, we represent these as functions of the QP mode operators using~\eqref{eq:Gutzwiller_Ops} in combination with the inversion of~\eqref{eq:QP_modeop1}. %and~\eqref{eq:QP_modeop2}. 
Utilizing the commutation relations %~\eqref{eq:GWfluc_commutk_red} %and~\eqref{eq:GWfluc_commutk_zero} 
for $R \rightarrow 0$ we can calculate spectral properties of the QP ground state $| \psi_{\textrm{QP}} \rangle$ implicitly defined via $\beta_{\mathbf{k},\gamma} | \psi_{\textrm{QP}} \rangle = 0$ for all $\mathbf{k}$ and $\gamma$. In the following we are especially interested in two types of spectral functions. Omitting the index $\sigma$, using $\tilde{H} = \hat{H} - E_0$ where $E_0$ is the energy of the QP ground state and introducing the notation $\langle \cdot \rangle_{\textrm{QP}} \equiv \langle \psi_{\textrm{QP}} | \cdot | \psi_{\textrm{QP}} \rangle $, the momentum-resolved spectral function $\mathcal{A}(\mathbf{k},\omega) = - { \textrm{sgn}(\omega)} \textrm{Im} \left[ \sum_{j,j'} e^{-i \mathbf{k} \cdot ( \mathbf{r}_j - \mathbf{r}_{j'})}  G_{jj'}(\omega) \right]/{ L \pi}  =  \langle \theta(\omega)  \tilde{b}_{\mathbf{k}} \delta( \tilde{H} - \omega) \tilde{b}^{\dag}_{\mathbf{k}} - \theta(-\omega) \tilde{b}^{\dag}_{\mathbf{k}} \delta(\tilde{H} + \omega) \tilde{b}_{\mathbf{k}} \rangle _{\textrm{QP}}$ is defined via the single-particle lattice Green's function $G_{ij}(\omega)$. Secondly, we consider the dynamic structure factor $S(\mathbf{k},\omega) =  \langle \tilde{n}_{\mathbf{k}} \delta(\tilde{H} - \omega) \tilde{n}_{\mathbf{k}} \rangle _{\textrm{QP}}$. % and the leading order local response $\delta O_i(\mathbf{k},\omega(\mathbf{k},\gamma))$ of a local operator $\hat{O}_{\sigma,i}$ obtained for a weak coherent excitation $| z,\mathbf{k},\gamma \rangle = \exp(-|z|^2/2) \exp(z \beta^{\dag}_{\mathbf{k},\gamma})|\psi_{\textrm{QP}} \rangle$. 
%Omitting the index $\sigma$, using $\tilde{H} = \hat{H} - E_0$ where $E_0$ is the energy of the QP ground state and $\langle \cdot \rangle_{\textrm{QP}} \equiv \langle \psi_{\textrm{QP}} | \cdot | \psi_{\textrm{QP}} \rangle $, one finds

%\begin{align}
%& \mathcal{A}(\mathbf{k},\omega) = - \frac{ \textrm{sgn}(\omega)}{ L \pi} \textrm{Im} \left[ \sum_{j,j'} e^{-i \mathbf{k} \cdot ( \mathbf{r}_j - \mathbf{r}_{j'})}  G_{jj'}(\omega) \right] \\ 
%& =  \langle \theta(\omega)  \tilde{b}_{\mathbf{k}} \delta( \tilde{H} - \omega) \tilde{b}^{\dag}_{\mathbf{k}} - \theta(-\omega) \tilde{b}^{\dag}_{\mathbf{k}} \delta(\tilde{H} + \omega) \tilde{b}_{\mathbf{k}} \rangle _{\textrm{QP}}, \nonumber \\
%& S(\mathbf{k},\omega) =  \langle \tilde{n}_{\mathbf{k}} \delta(\tilde{H} - \omega) \tilde{n}_{\mathbf{k}} \rangle _{\textrm{QP}} .
% & \delta O_i(\mathbf{k},\omega) = \langle z^* \beta_{\mathbf{k},\gamma} \hat{O}_{\sigma,i} + z \hat{O}_{\sigma,i} \beta_{\mathbf{k},\gamma}^{\dag} \rangle _{\textrm{QP}} / |z|
%\end{align}

\paragraph*{Results} \label{sec:ONdress_finiteSS}

In the presence of blackbody radiation and spontaneous decay, both driving transitions to nearby Rydberg $p$-states for the $s$-state Rydberg excitations considered here, there exists a dissipative channel generating a strong global loss of coherence and atoms due to incoherent $s$-$p$-dipole scattering, referred to as F\"orster processes  \cite{Gunter2013,Goldschmidt2016,Aman2016,Zeiher2016}. The important figure of merit in this context is the time $\tau$ until creation of the first contaminant $p$-state. Due to the underlying strong dipole interactions this is a global process, therefore one has to consider the creation of such a state anywhere in the system, implying an integration of the Rydberg fraction over all sites,
%\begin{align}
$\tau^{-1} = b \Gamma_0 \sum_i \langle \hat{n}^e_i \rangle$. %\label{eq:AV_timescale}
%\end{align}
The term $b$ denotes the branching ratio of the decay into detrimental states, with a typical value of $ b \gtrsim 20 \%$ at room temperature, and $\Gamma_0$ is the full natural decay rate, for example, $\Gamma_0^{-1} = \tau_0 \gtrsim 30 \mu$s for the $36S$ state of $^{87}$Rb \cite{Beterov2009b,Boulier2017}. While this expression diverges with the system size for any nonzero Rydberg concentration, it is possible to limit its value in a finite system. %In the far-detuned regime $n_{ei} = (\Omega / |2 \Delta|)^2$, so $n_{ei} = 1\%$ for $|\Delta| / \Omega = 5$.
Even at near-resonant excitation of Rydberg states we obtain a local Rydberg fraction of $\approx 0.1\%$  in the bulk of the system, as shown in Fig.\ref{fig:trapped_SS}. Since this value is achieved via a many-body induced MF shift
$
\Delta^{i}_{\textrm{shift}} = - V \sum_{j \neq i} {n^e_j}{\left| \mathbf{i}-\mathbf{j} \right|^{-6}}
$
of the detuning at site $i$, thus blocking nearby Rydberg exciations, one has to be careful at the boundary, which in our case is given by the intensity beam-profile of the Rabi laser. The number of neighboring Rydberg excitations is significantly reduced for a sufficiently sharp (beam-)edge, resulting in a reduced  $\Delta^{i}_{\textrm{shift}}$. Thus the effective Rabi process becomes increasingly near-resonant at the edge, leading to a strong increase in the Rydberg fraction, easily $50$ times the bulk value. Therefore the excitations forming at the edge  \cite{Schauss2012,Schauss2015,Vermersch2015} quickly induce the avalanche on a time scale which is well approximated by the bare lifetime. %$\tau = \mathcal{O} (\tau_0)$. 
The observation of itinerant physics in this limit is thus unlikely without suppression of the detrimental transitions.

%\begin{table}[h]
%\centering
%\captionsetup{width=0.95\textwidth}
%\begin{tabular}{c c c c c c c c c}
%$U_{gg}$	&$U_{ge}$ &$U_{ee}$&$J$&$\mu$&$\Delta$&$\Omega$&$V$ \\
%\hline
%$3\times 10^{-3}$	&$5$ &$100$&$3\times 10^{-4}$&$3\times 10^{-4}$&$-0.2$&$0.2$&$10^4$
%\end{tabular}
%\caption{
%Model parameters used for the finite lattice system~\eqref{eq:full-hamil}. Using $\hbar = 1$ all values are given in units of MHz.
%}
%\label{tab:trapped_params}
%\end{table}

In our simulations we consider a periodic system of $21 \times 21$ sites and parameters as given in Fig.~\ref{fig:trapped_SS}. It is driven by an inhomogeneous Rabi laser, determining the geometry of the system, which has a beam-waist described by %an exponent $\gamma$
%\begin{equation} \label{eq:beam-waist}
$\Omega (d) = \Omega \exp \left[-  \left( {d}/{\kappa} \right)^{\gamma} \right]$, 
%\end{equation}
where $d$ is the distance from the center of the beam and $\kappa = 8.5 a$. For $\gamma = 2$ one obtains the common Gaussian beam-waist, while it becomes increasingly box-like as $\gamma \rightarrow \infty$, reminiscent of optical box potentials \cite{Gaunt2013}. %Intermediate shapes can be achieved using spatial light modulators. 
The beam-waist separates the system in up to three parts: the central region with approximately constant Rabi frequency $\Omega(d) \approx \Omega$, the edge with vanishing $\Omega(d) \approx 0$, which can be considered as the central region of a system without Rydberg excitations, and the crossover region. %, which can be considered as a soft boundary.
We analyze the three cases $\gamma =  2,4,\infty $.

\begin{figure}[h]
 \centering
% \captionsetup{width=0.95\textwidth}
 \includegraphics[width=0.98\columnwidth]{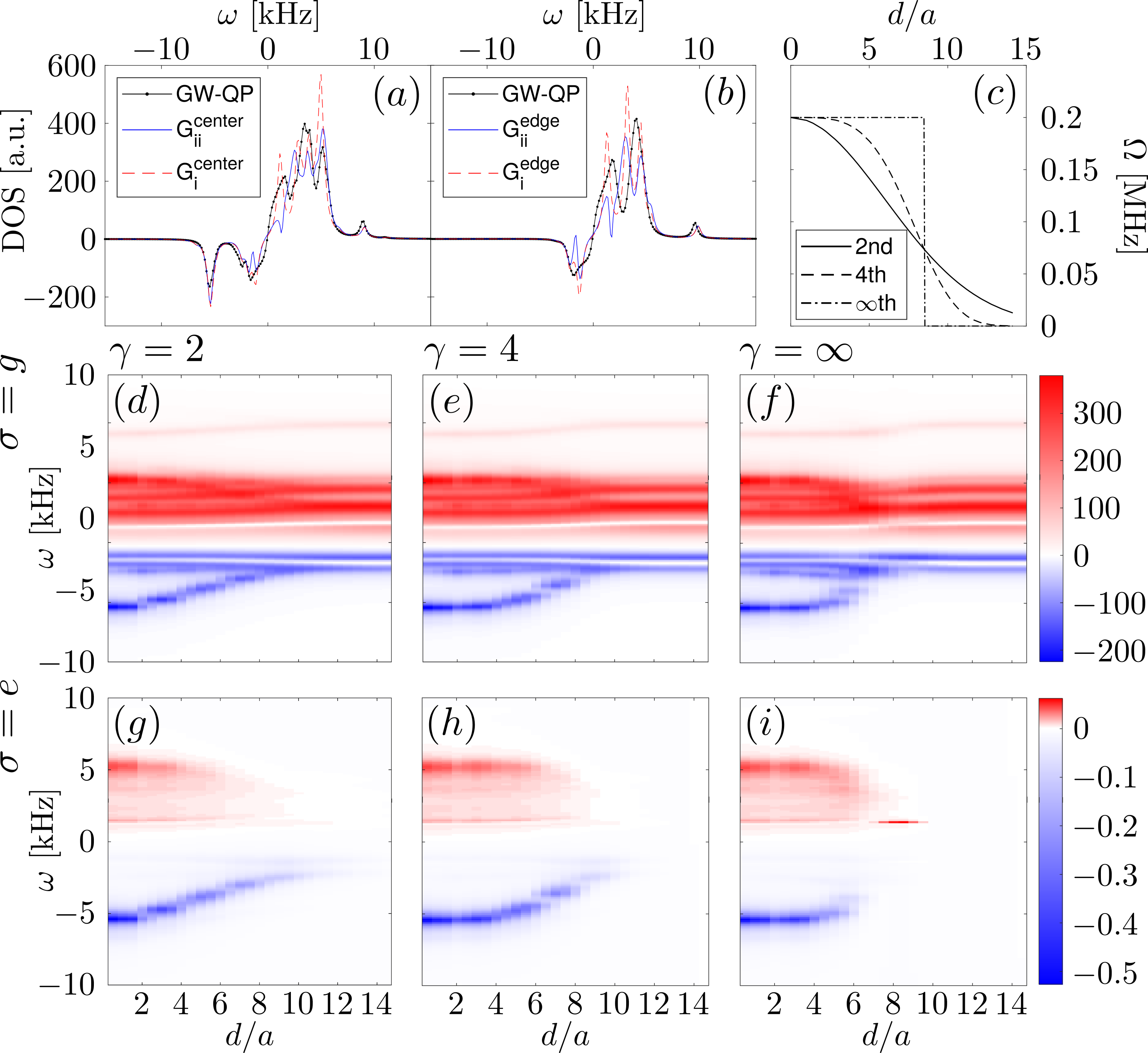}
 \caption{Spectral properties of finite Rydberg supersolids.  $(a,b)$ Comparison of the $\sigma = g$ l-DOS in real-space, from $G_{ii}(\omega)$ and $G_{i}(\omega)$ (see text), with the QP result of a $2^7 \times 2^7$ periodic lattice with fixed $\Omega$. Spectra correspond to $\gamma = 4$ and $d/a = 0.5,14$ $(a),(b)$ respectively. $(c)$ Considered radial shapes of the Rabi-laser beam, with $\kappa = 8.5 a$. $(d-i)$ Average l-DOS (from $G_{ii}(\omega)$) of the indicated component $\sigma$ as a function of the distance $d$ from the beam center for $\gamma = 2,4,\infty$.}
 \label{fig:trapped_specs}
\end{figure}

The idea behind this setup is to counteract the overshoot of the Rydberg fraction at a hard edge, such that we can obtain an overall more smooth distribution of the Rydberg fraction close to the bulk value, as shown in Fig.~\ref{fig:trapped_SS}. Using real-space extended bosonic dynamical mean-field theory (RB-DMFT) \cite{Byczuk2008,Hubener2009,Snoek2013,Geissler2017,Li2018}, we find a ground-state consisting of a 4-site-supersolid at the center of the beam with a Rydberg concentration $n_{ei} < 1.5 \times 10^{-3}$ and a homogeneous superfluid outside the beam (see Fig.~\ref{fig:trapped_SS}). Such a low Rydberg fraction, which is even below the Rydberg population of a typical off-resonant dressing scheme \cite{Zeiher2017}, results from the many-body blockade desribed above. We note that this implies a pairwise van der Waals interaction energy of less then $10^{-3}$MHz which is below the energy scale of $U_{gg}$, thus \textit{a posteriori} validating the single band assumption underlying~\eqref{eq:full-hamil}. For a sufficiently soft beam-edge we find a very low total Rydberg number: $\sum_i n_{ei} =  0.1024, 0.1068, 0.5722$ for $\gamma = 2,4,\infty$ respectively. This implies an avalanche timescale $\tau \approx 1.4 \, $ms for $\gamma = 4$, which is of the same order of magnitude as the tunneling time, so that we expect this state to be observable in experiments, especially with the additional aid of post-selection \cite{Zeiher2016} or a cryogenic environment \cite{Boulier2017}.

As the real-space results naturally contain the crossover between a homogeneous superfluid and the supersolid, we analyze the spectral transition between the two phases (see Fig.~\ref{fig:trapped_specs}). The radial dependence of the local density of states (l-DOS) obtained via RB-DMFT is averaged over rings with a width of two lattice sites (diameter of a unit cell). RB-DMFT is a non-perturbative method with which we self-consistently obtain the diagonal elements of the interacting lattice Green's function $G_{ii}(i \omega_n) \equiv G_{i}(i \omega_n)$ in Matsubara frequencies, where the right-hand side is the interacting Green's function of individual effective local Anderson impurity models obtained by tracing the remaining system \cite{Hubener2009,Anders2010,Snoek2013,Geissler2017}. From both representations we calculate $\mathcal{A} (\mathbf{r}_i,\omega) = - \textrm{Im} \left[ G_{ii / i}( \omega) \right] / \pi $ via analytic continuation. The most remarkable features we find at the crossover are the separation of a very narrow gapped hole mode driven by the excited component (see Fig.~\ref{fig:trapped_specs}$(g-i)$), as well as a broadening of the particle modes. Furthermore, we find localized modes at the boundary of the $\gamma = \infty$ beam in Fig.~\ref{fig:trapped_specs}(i) which we associate with the Rydberg excitations localized at the beam-edge. 

A comparison to QP results reveals the agreement between both methods, which all consider a Lorentzian broadening of $0.3$kHz. The observed discrepancies stem from the inherently discretized nature of the DMFT spectrum %due to the discrete nature of the considered impurity models, 
in combination with the incapability of DMFT to properly describe long-wavelength Goldstone modes \cite{Panas2015,Panas2017}. %always appearing gapped in DMFT. The reason for this is the limitation of DMFT to the lowest order non-local correlations. 
While all $\sigma = g$ l-DOS fulfill the sum rule, $\int A(\mathbf{k},\omega) d\omega = 1$, to within $3 \%$, where the QP result is the closest (typically $< 10^{-4}$, signifying the applicability of the QP theory) and the result from the lattice Green's function deviates the most, all $\sigma = e$ l-DOS closely equate to zero due to the quantum Zeno assumption. % The remaining deviations primarily stem from the numerically neccesary Hilbert space truncations which is N=4 in DMFT and N=6 in the QP case.

\begin{figure}[h]
 \centering
% \captionsetup{width=0.95\textwidth}
 \includegraphics[width=0.98\columnwidth]{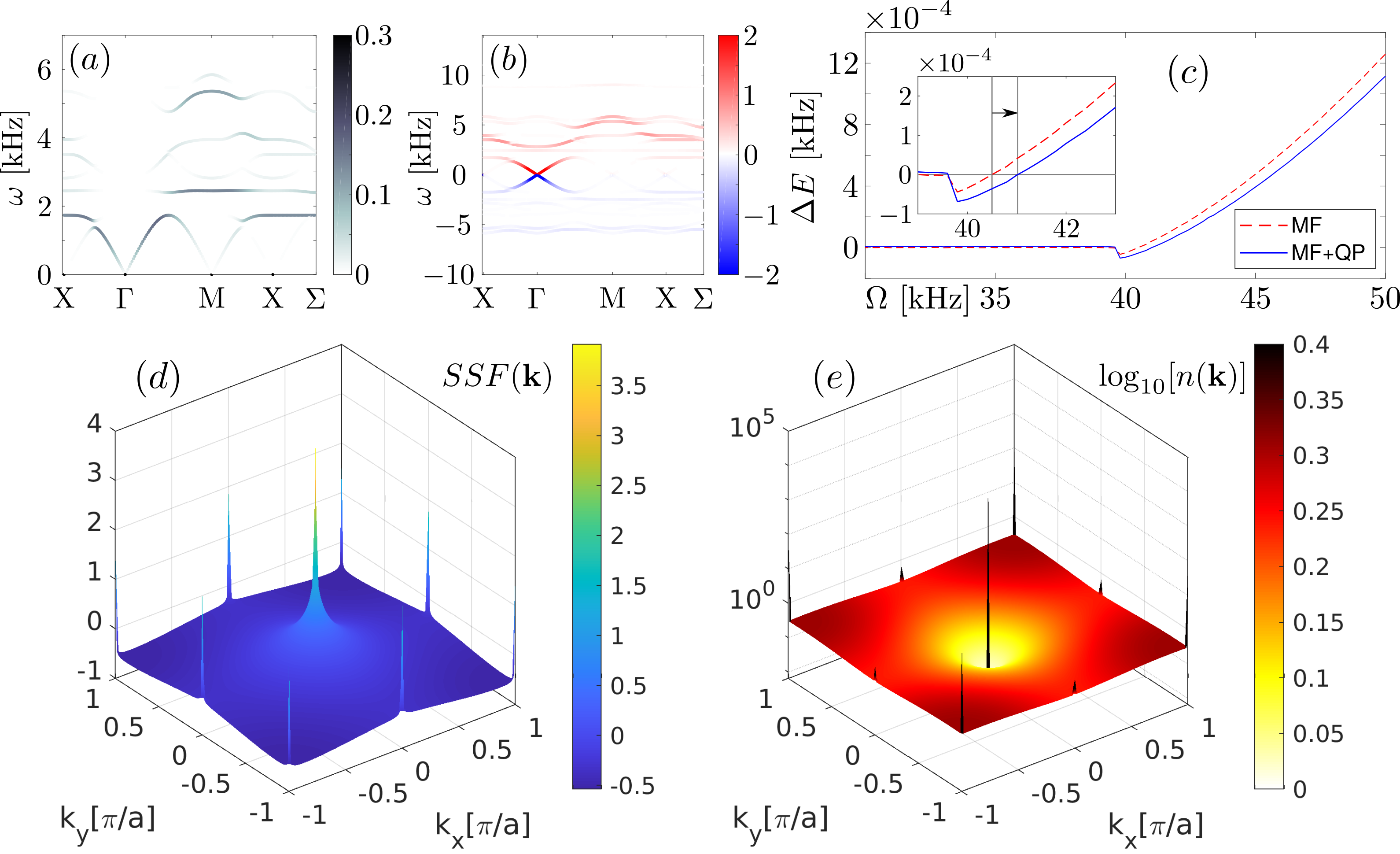}
 \caption{Spectral properties of the ground state species. 
 %$(a)-(c)$ From left to right: average local responses $\textrm{Re}(\delta b_g)$, $\textrm{Im}(\delta b_g)$ and $\delta n_g$. 
 $(a)$ Dynamic structure factor and $(b)$ single particle spectral density along high symmetry points in the first Brillouin zone of the optical lattice, $(d)$ static structure factor and $(e)$ momentum distribution. $(c)$ $\Delta E$ introduced in the text on the MF and MF+QP level. The arrow in the inset indicates the QP renormalization of the checkerboard to 4-site unit cell transition (grey lines are guides to the eye).}
 \label{fig:QP_specs}
\end{figure}

We obtain more detailed results for the supersolid phase via the QP method, assuming a system of $2^7 \times 2^7$ sites and a homogeneous Rabi frequency. As the MF ground-state has a square 2-by-2 unit cell, the bands are back-folded so that four reduced Brillouin zones ($\textrm{BZ}'$) lie within the first Brillouin zone ($1.\textrm{BZ}$) of the underlying lattice. %This is seen best in the local responses characterizing the QP modes, see Fig.~\ref{fig:QP_specs}. While we find only one gapless band with four Nambu-Goldstone cones in the $1.\textrm{BZ}$, associated with the condensate fraction and its spatial variation (see Fig.~\ref{fig:QP_specs}$a,b,d$ and $g$), the gapped amplitude modes are folded back multiple times and exhibit avoided crossings at degenerate points. The amplitude modes are furthermore mixed with density fluctuations and where density fluctuations are small, the condensate fraction fluctuates at a fixed particle density (see Fig.~\ref{fig:QP_specs}$a,b,c$).
In experiment, the $g$ component is more accessible to direct observation and for its spectral function and dynamic structure factor (see Fig.~\ref{fig:QP_specs}$(a),(b)$) we observe that only the central one ($|\mathbf{k}| \approx 0$) of the possible four ungapped Goldstone cones in the $1.\textrm{BZ}$ yields a significant contribution. For all other momenta, only gapped modes contribute, revealing different parts of the excitation spectrum in each quantity. As expected from the supersolid state, we find a total of four peaks in the static structure factor $SSF(\mathbf{k}) = \int S(\mathbf{k},\omega) d \omega$ and the momentum distribution $n(\mathbf{k}) = \langle \tilde{n}_{\mathbf{k}} \rangle_{\textrm{QP}}$ (see Fig.~\ref{fig:QP_specs}$(d),(e)$) at the high symmetry points $\Gamma,M$ and $X$ reflecting the spatial symmetry breaking of the SS.

Due to the inhomogeneity caused by the beam-waist we observe a direct transition from the 2-by-2 unit cell supersolid to a homogeneous superfluid within RB-DMFT. But using the QP method we further find an intermediate checkerboard (CB) supersolid, whose transition to the 2-by-2 supersolid is shifted in relation to the MF result when including the correction $\delta E_{\textrm{QP}}$. Considering the energy differences $\Delta E_{\textrm{MF(+QP)}} = E_{\textrm{MF(+QP)}}^{\textrm{CB}} - E_{\textrm{MF(+QP)}}^{\textrm{2-by-2}}$ we quantify this correction in Fig.~\ref{fig:QP_specs}$(c)$.

\paragraph*{Conclusion}

Exploiting the inhomogeneity of the Rabi laser beam-waist, we show that the Rydberg fraction of a driven lattice gas can be reduced significantly, even if the Rydberg excitation is near-resonant. As a result, it is possible to obtain an extended supersolid consisting of roughly 50 2-by-2 unit cells while having only a small fraction of a Rydberg excitation present in the whole system. The corresponding avalanche time scale is therefore vastly enhanced, thus paving the way for realizing a supersolid state even at near-resonant excitation where the blockade radius vanishes. In practice, one can also reduce the number of unit cells using an even narrower beam-waist to further reduce the total Ryberg fraction or use higher Rydberg levels to enhance the many-body blockade. Furthermore, a promising starting point before the switch-on of the Rabi laser would be a Mott state at unit filling or a low density condensate, as to suppress loss due to the mentioned molecule formation. Additionally, we have analyzed the spectral properties of the crossover between the supersolid and the superfluid bulk in terms of the l-DOS obtained via RB-DMFT as well as the momentum resolved spectral properties in the supersolid bulk obtained via a generalized QP method including all higher QP modes. The latter procedure also determines a correction of the static MF energies resulting in a correction of the MF phase transition.

\begin{acknowledgments}

We would like to thank C. Gro{\ss}, S. Hollerith and H. Weimer for insightful discussions. Support by the Deutsche Forschungsgemeinschaft via DFG SPP 1929 GiRyd and the high-performance computing center LOEWE-CSC is gratefully acknowledged.

\end{acknowledgments}

\appendix
\numberwithin{equation}{section}
\section[A]{Deviation from bosonic behavior}\label{sec:Commutator_deviation}

In the main part we introduce the operator $R^{(n,m)}_{\mathbf{k},\mathbf{k}'}(s)$ to describe the deviation from bosonic behavior for the Gutzwiller fluctuation operators $\hat{\sigma}_{l,s}^{(i)^{\dag}}$ and $\hat{\sigma}_{l,s}^{(i)}$. Its precise form is given by the expression

\begin{widetext}
\begin{align}
R^{(n,m)}_{\mathbf{k},\mathbf{k}'}(s) =& \sum_{l} e^{i(\mathbf{k}-\mathbf{k}')\cdot (\mathbf{r}_l + \mathbf{r}_s)} \left( \hat{\sigma}_{l,s}^{{(n)}^{\dag}} \hat{\sigma}_{l,s}^{(m)} + \delta_{n,m} \sum_{m'>0} \hat{\sigma}_{l,s}^{{(m')}^{\dag}} \hat{\sigma}_{l,s}^{(m')} \right) \\
=& \sum_{\mathbf{k}_1 \in 1.\textrm{BZ}'} \left(\tilde{\sigma}^{(n)^{\dag}}_{ \lfloor \mathbf{k-k'+k}_1 \rfloor,s} \tilde{\sigma}^{(m)}_{\mathbf{k}_1,s} + \delta_{n,m} \sum_{m'>0} \tilde{\sigma}^{(m')^{\dag}}_{ \lfloor \mathbf{k-k'+k}_1 \rfloor,s} \tilde{\sigma}^{(m')}_{\mathbf{k}_1,s} \right). \label{eq:R_of_k}
\end{align}
\end{widetext}
The notation $\lfloor \cdot \rfloor$ describes the back folding of $\mathbf{k}$ to the first Brillouin zone ($1.\textrm{BZ}$) introduced due to equivalence relations between the quasimomentum space operators $\tilde{\sigma}^{(n)^{\dag}}_{\mathbf{k},s}$ and $\tilde{\sigma}^{(n)}_{\mathbf{k},s}$. Back folding is achieved by adding a suitable reciprocal lattice vector $\mathbf{G}$, such that $\lfloor \mathbf{k} \rfloor = \mathbf{k+G}$ results in a quasimomentum vector inside the $1.\textrm{BZ}$. The set of possible vectors $\mathbf{G}$ is implicitly defined as all vectors fulfilling the relation $\mathbf{G} \cdot \mathbf{r}_{i} = 2 \pi n$ where $n \in \mathbb{Z}$. For states with a reduced lattice symmetry only the inequivalent quasimomenta inside the reduced Brillouin zone ($1.\textrm{BZ}'$) are considered in the summation~\eqref{eq:R_of_k}, which is given via the reduced reciprocal lattice vectors $\mathbf{G}_r$ implicitly defined by the relation $\mathbf{G}_r \cdot \mathbf{r}_{l} = 2 \pi n$ for all $n \in \mathbb{Z}$.

Thus we can see that the deviation from bosonic commutation relations is on the order of the density of Gutzwiller fluctuations, especially notable for $\mathbf{k} = \mathbf{k}'$. As this factor is scaled by the system size via the prefactor $L_c^{-1}$, the fluctuation operators $\tilde{\sigma}^{(i)}_{\mathbf{k}}$ are also approximately bosonic in the limit of a small density of occupied fluctuations. Therefore we may consider $\langle R^{(i,j)}_{\mathbf{k},\mathbf{k}'}(s) \rangle / L_c$ as a set of control parameters, the upper bound of which serves as the figure of merit for the validity of the quasiparticle method in the main part. 

\section[B]{The quasiparticle Hamiltonian $\mathcal{H}^{(2)}_{\textrm{QP}}$}\label{sec:qp-Hamil-diag}

We use the partial Fourier-transformed Gutzwiller operators 
\begin{align}
\tilde{\sigma}^{(n)}_{\mathbf{k},s} = L_c^{-1/2} \sum_{l} e^{-i\mathbf{k} \cdot (\mathbf{r}_l+\mathbf{r}_s)} \hat{\sigma}^{(n)}_{l,s}, \\ \tilde{\sigma}^{(n)^{\dag}}_{\mathbf{k},s} = L_c^{-1/2} \sum_{l} e^{i\mathbf{k} \cdot (\mathbf{r}_l+\mathbf{r}_s)} \hat{\sigma}^{(n)^{\dag}}_{l,s},
\end{align}
as we assume an inhomogeneous many-body ground-state consisting of a periodic distribution of finite-sized unit cells. Thus a general position vector $\mathbf{r}_i = \mathbf{r}_l + \mathbf{r}_s$ separates into a Bravais lattice contribution of unit cells $l$ and the relative position of distinct sites $s$ in each unit cell. The partial Fourier transform of the vectors $ \pmb{\sigma} = \left( \sigma_{1}^{(1)} , \ldots  , \sigma_{L}^{(N)} \right)^{\textrm{T}}$ and $ \pmb{\sigma}^{\dag} = \left( \sigma_{1}^{(1)^{\dag}} , \ldots  , \sigma_{L}^{(N)^{\dag}} \right)^{\textrm{T}}$ defines the corresponding vectors $\tilde{\pmb{\sigma}}$ and $\tilde{\pmb{\sigma}}^{\dag}$. Using these and taking the approximation $R \rightarrow 0$ when using the commutation relation (4) discussed in the main part yields the following form of the second order term $\mathcal{H}^{(2)}$:

\begin{align}
\mathcal{H}^{(2)} \approx \mathcal{H}^{(2)}_{\textrm{QP}} = \frac{1}{2} \begin{pmatrix}
\tilde{\pmb{\sigma}} \\ \tilde{\pmb{\sigma}}^{\dag}
\end{pmatrix} ^{\dag} \tilde{\mathcal{H}}_{\textrm{QP}} \begin{pmatrix}
\tilde{\pmb{\sigma}} \\ \tilde{\pmb{\sigma}}^{\dag}
\end{pmatrix} -\frac{1}{2} \textrm{Tr}(h), \label{eq:H2_in_quasimomentum} \\
\textrm{with} \quad
\tilde{\mathcal{H}}_{\textrm{QP}}(\mathbf{k}) = 
\begin{pmatrix}
\tilde{h}(\mathbf{k}) & \tilde{\Delta}(\mathbf{k}) \\
\tilde{\Delta}(-\mathbf{k})^* & \tilde{h}(-\mathbf{k})^*  
\end{pmatrix}.
\label{eq:operatorQPH}
\end{align}
Within this approximation the introduced Hamiltonian matrix $\mathcal{H}_{\textrm{QP}}$ is block-diagonal with blocks of the size $2 N_c N \times 2 N_c N$, as $\mathbf{k} \leftrightarrow \mathbf{-k}$ sectors are coupled. Its individual matrix elements are given in terms of $\prescript{}{i}{\langle} n | \hat{  b}_{\sigma i} | m \rangle_i$ and $\prescript{}{i}{\langle} n | \hat{  n}_{\sigma i} | m \rangle_i$ matrix elements in the local Gutzwiller bases, while equivalent sites $i$ are mapped onto their representative $s(i)$. For the extended two-component Bose-Hubbard model with long-range interactions (1) considered in the main part the explicit matrix entries of the individual $\mathbf{k}$-blocks are given by

\begin{widetext}
\begin{align}
\tilde{h}_{(n,s),(m,s')}(\mathbf{k}) &= - \sum_{\sigma} F_{n,0,0,m}^{\sigma(s,s')} J_{\sigma}^{(s,s')}(\mathbf{k}) + {N}_{n,0}^{e,(s)} {N}_{0,m}^{e,(s')} V^{(s,s')}(\mathbf{k}) + \delta_{s,s'} \delta_{n,m} E_n^{(s)}, \label{eq:QPmatrix_h} \\
\tilde{\Delta}_{(n,s),(m,s')}(\mathbf{k}) &= - \sum_{\sigma} F_{n,m,0,0}^{\sigma(s,s')} J_{\sigma}^{(s,s')}(\mathbf{k}) + {N}_{n,0}^{e,(s)} {N}_{m,0}^{e,(s')} V^{(s,s')}(\mathbf{k}). \label{eq:QPmatrix_Delta}
\end{align}
\end{widetext}
These expressions are given in terms of the restricted Fourier transforms of the hopping and interaction matrices $J_{\sigma}^{(i,j)} = J \, \forall \, \{\langle i,j \rangle\}$ and $V^{(i,j)} = V / d_{ij}^{6}$ respectively. While $E_n^{(s)}$ are the eigenenergies of the $n$th Gutzwiller excited state for each representative site $s$, the remaining terms are the matrix elements of the non-local products of local operators

\begin{align}
{N}^{\sigma,(s)}_{n,m} &= \prescript{}{s}{\left\langle n \right|} \hat{n}_{\sigma s} \left| m \right\rangle_{s} - n_{s\sigma} \delta_{n,m}, \\
{F}^{\sigma,(s,s')}_{n_1,n_2,m_1,m_2} &= {B}^{\sigma,(s)*}_{m_1,n_1} {B}^{\sigma,(s')}_{n_2,m_2} + {B}^{\sigma,(s)}_{n_1,m_1} {B}^{\sigma,(s')*}_{m_2,n_2} \\ \textrm{with} \quad {B}^{\sigma,(s)}_{n,m} &= \prescript{}{s}{\left\langle n \right|} \hat{b}_{\sigma s} \left| m \right\rangle_{s} - \phi_{s \sigma} \delta_{n,m}, \nonumber
\end{align}
where $n_{s \sigma}$ and $\phi_{s \sigma}$ have been defined as the self-consistent mean-field values obtained for the many-body ground-state. Note that this identification is the reason for the absence of any first order term in the Gutzwiller fluctuation representation of $\Delta \left( \hat{\delta}^2 \right)$.

\subsection*{Diagonalization of $\mathcal{H}^{(2)}_{\textrm{QP}}$}

In order to preserve the bosonic structure of the operators the diagonalization of~\eqref{eq:H2_in_quasimomentum} has to be performed on the symplectic space, namely by diagonalizing $\Sigma \mathcal{H}^{(2)}_{\textrm{QP}}$, where $\Sigma = \textrm{diag} (\mathbb{1}_{NL} , - \mathbb{1}_{NL})$. This yields the representation of $\mathcal{H}^{(2)}$ in terms of QP mode operators which are defined as~\eqref{eq:QP_modeop1} and

\begin{align}
%\beta_{\mathbf{k},\gamma} &\equiv \mathbf{x}^{(\mathbf{k},\gamma)^{\dag}} \Sigma \begin{pmatrix} \tilde{\boldsymbol{\sigma}} \\ \tilde{\boldsymbol{\sigma}}^{\dag} \end{pmatrix} \equiv \mathbf{u}^{(\mathbf{k},\gamma)^{\dag}}\tilde{\boldsymbol{\sigma}} + \mathbf{v}^{(\mathbf{k},\gamma)^{\dag}} \tilde{\boldsymbol{\sigma}}^{\dag}. \label{eq:QP_modeop1} \\
\beta_{\mathbf{k},\gamma}^{\dag} &\equiv -\mathbf{y}^{(\mathbf{k},\gamma)^{\dag}} \Sigma \begin{pmatrix} \tilde{\boldsymbol{\sigma}} \\ \tilde{\boldsymbol{\sigma}}^{\dag} \end{pmatrix} \equiv \mathbf{v}^{(\mathbf{k},\gamma)^{T}}\tilde{\boldsymbol{\sigma}} + \mathbf{u}^{(\mathbf{k},\gamma)^{T}} \tilde{\boldsymbol{\sigma}}^{\dag}. \label{eq:QP_modeop2}
\end{align}
They are given by the eigenvectors of the eigenvalue equations $\Sigma \tilde{\mathcal{H}}_{\textrm{QP}} ^{(2)} \mathbf{x}^{(\mathbf{k},\gamma)} = \omega_{\mathbf{k},\gamma} \mathbf{x}^{(\mathbf{k},\gamma)}$ and $\Sigma \tilde{\mathcal{H}}_{\textrm{QP}} ^{(2)} \mathbf{y}^{(\mathbf{k},\gamma)} = -\omega_{\mathbf{k},\gamma}^* \mathbf{y}^{(\mathbf{k},\gamma)}$. Thus all  
QP frequencies $\omega_{\mathbf{k},\gamma}$ appear in pairs and those with a non-zero imaginary part represent unstable QP modes that are commonly only encountered for MF states far from the ground-state.

In the presence of a condensate fraction the $\mathbf{k} = 0$ QP frequencies in the lowest band $\gamma=1$ vanish, so there is a degenerate subspace. Then the eigenvalue equation reduces to $\Sigma \tilde{\mathcal{H}}_{\textrm{QP}}(\mathbf{k} = 0) \mathbf{p} = 0$ which is solved by an eigenvector of the form $\mathbf{p} = (\mathbf{u}^{(0)} , -\mathbf{u}^{(0)^*})^T$. In order to complete the representation of this subspace one further has to introduce a second vector $\mathbf{q}$ within it, which is implicitly defined via $\Sigma \tilde{\mathcal{H}}_{\textrm{QP}}(\mathbf{k} = 0) \mathbf{q} = -i \mathbf{p} / \tilde{m}$, where $\tilde{m}$ is a mass-like scalar. Therefore we find two different operators taking the places of the Bogoliubov-like operators~\eqref{eq:QP_modeop1} and~\eqref{eq:QP_modeop2} for $(\mathbf{k} = 0,\gamma = 1)$ (which have been discussed in further detail in \cite{Bissbort2012}):

\begin{align}
\mathcal{P} &\equiv \mathbf{p}^{\dag} \Sigma \begin{pmatrix} \tilde{\boldsymbol{\sigma}} \\ \tilde{\boldsymbol{\sigma}}^{\dag} \end{pmatrix} = \begin{pmatrix} \mathbf{u}^{(0)} \\ -\mathbf{u}^{(0)^*} \end{pmatrix}^{\dag} \Sigma \begin{pmatrix} \tilde{\boldsymbol{\sigma}} \\ \tilde{\boldsymbol{\sigma}}^{\dag} \end{pmatrix},  \label{eq:QP_modeop_P} \\
\mathcal{Q} &\equiv -\mathbf{q}^{\dag} \Sigma \begin{pmatrix} \tilde{\boldsymbol{\sigma}} \\ \tilde{\boldsymbol{\sigma}}^{\dag} \end{pmatrix} \equiv i \begin{pmatrix} \mathbf{v}^{(0)} \\ -\mathbf{v}^{(0)^*} \end{pmatrix}^{\dag} \Sigma \begin{pmatrix} \tilde{\boldsymbol{\sigma}} \\ \tilde{\boldsymbol{\sigma}}^{\dag} \end{pmatrix}.  \label{eq:QP_modeop_Q}
\end{align}

Assuming exactly bosonic commutation relations for the Gutzwiller fluctuation operators and thus also for the QP mode operators, the second order quasiparticle term $\mathcal{H}^{(2)}_{}$ generally has the approximate form

\begin{align}
\mathcal{H}^{(2)}_{} \approx \tilde{\sum_{\mathbf{k},\gamma}} \omega_{\mathbf{k},\gamma} \beta_{\mathbf{k},\gamma}^{\dag} \beta_{\mathbf{k},\gamma} + \frac{\mathcal{P}^2}{2\tilde{m}} + \left( \tilde{\sum_{\mathbf{k},\gamma}} \frac{\omega_{\mathbf{k},\gamma}}{2} - \frac{\textrm{Tr}(h)}{2} \right). \label{eq:HQP_diag_fullP}
\end{align}
This representation is given in terms of the generalized Bogoliubov creation (annihilation) operators $\beta_{\mathbf{k},\gamma}^{\dag}$ ($ \beta_{\mathbf{k},\gamma}$) where the notation $\tilde{\sum}_{\mathbf{k} ,\gamma}$ represents the fact that the $(\mathbf{k} = 0,\gamma =1)$ term in the sum is to be replaced by $\mathcal{P}$ whenever a condensate is present. Otherwise there is no $\mathcal{P}$ term. We note that $\mathcal{P}$ is a momentum-like operator which can be considered as the generator of translations in the global phase of the condensate mode \cite{Lewenstein1996}, so it represents the free motion of the condensate phase. From~\eqref{eq:HQP_diag_fullP} we can thus see that the quasiparticle ground-state is characterized by $\langle \psi_{\textrm{QP}} | \mathcal{P}^2 | \psi_{\textrm{QP}} \rangle = 0$ and we can use $\mathcal{P} | \psi_{\textrm{QP}} \rangle = 0$. Regarding the spectral properties discussed in the main part, consideration of $\mathcal{P}$ and $\mathcal{Q}$ only yields a vanishingly small correction at $\mathbf{k}=0$ and $\omega = 0$ in the thermodynamic limit, that even is self-canceling for $\mathcal{A}(\mathbf{k},\omega)$, so we may neglect both for our purposes.

By expressing the Hamiltonian with $\beta_{\mathbf{k},\gamma}^{\dag}$ and $ \beta_{\mathbf{k},\gamma}$ in normal order we find the scalar contribution proportional to $\tilde{\sum}_{\mathbf{k},\gamma} \omega_{\mathbf{k},\gamma}$. Note that the last two scalar terms in~\eqref{eq:HQP_diag_fullP} generate a shift of the total energy. While both contributions $\tilde{\sum}_{\mathbf{k},\gamma} \omega_{\mathbf{k},\gamma}$ and $\textrm{Tr}(h)$ would diverge individually in the limit of no truncation ($N \rightarrow \infty$), even in a finite system, in combination they yield a finite correction of the quasiparticle ground-state energy density. They effectively lower its value in relation to the Gutzwiller MF result due to the average shift of the mode energies in relation to the energies of the Gutzwiller excitations.

\section[C]{Order parameter fluctuations - band characterization}

\begin{figure}[h]
 \centering
% \captionsetup{width=0.95\textwidth}
 \includegraphics[width=0.99\columnwidth]{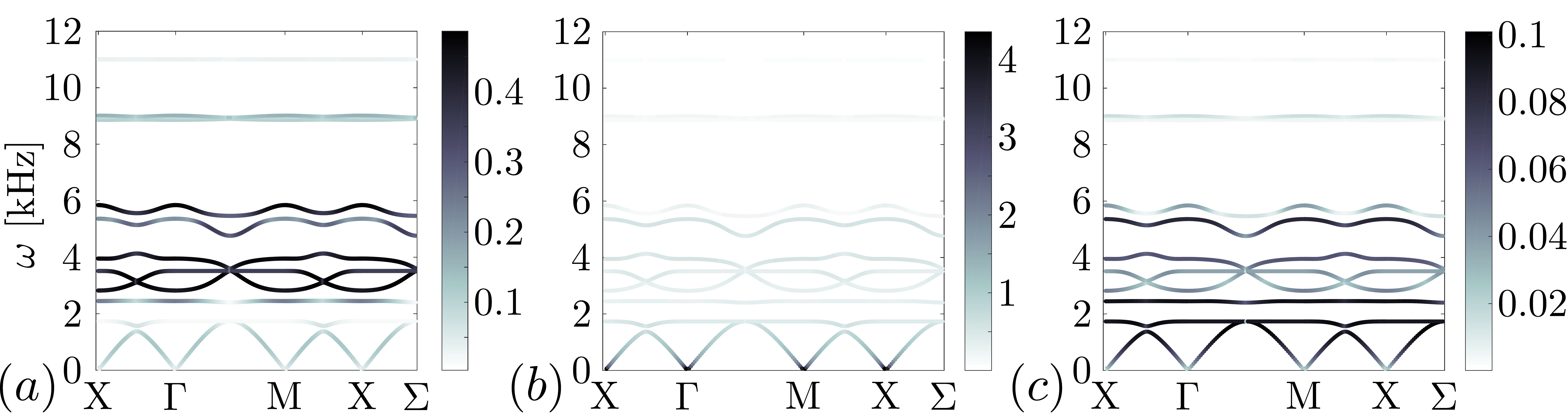}
 \caption{Quasiparticle responses of the 2-by-2 supersolid along the high symmetry points of the optical lattice: Amplitude response $\textrm{Re} (\delta b_{g})$ $(a)$ and phase response $\textrm{Im} (\delta b_{g})$ $(b)$ of the condensate order parameter, as well as the density response $\delta n_{g}$ $(c)$. Parameters are the same as used in the main part and $\Omega = 0.2\textrm{ MHz}$.}
 \label{fig:QP_response}
\end{figure}

In addition to the spectral properties discussed in the main part one can also consider the leading order local response $\delta O_i(\mathbf{k},\omega(\mathbf{k},\gamma))$ of a local operator $\hat{O}_{\sigma,i}$ obtained for a weak coherent excitation $| z,\mathbf{k},\gamma \rangle = \exp(-|z|^2/2) \exp(z \beta^{\dag}_{\mathbf{k},\gamma})|\psi_{\textrm{QP}} \rangle$ in order to characterize the QP modes:

\begin{align}
\delta O_i(\mathbf{k},\omega) = {}_{\textrm{QP}}\langle z^* \beta_{\mathbf{k},\gamma} \hat{O}_{\sigma,i} + z \hat{O}_{\sigma,i} \beta_{\mathbf{k},\gamma}^{\dag} \rangle _{\textrm{QP}} / |z|.
\end{align}
There are three relevant quantities to be considered here. The first two are the real and the imaginary parts of the local response in the annihilation operator $\hat{b}_{gi}$, which for a real-valued order parameter represent amplitude and phase fluctuations respectively. Lastly, the local response of the number operator $\hat{n}_{gi}$ reveals modes with pronounced density fluctuations. In Fig.~\ref{fig:QP_response} we show the lattice averages $\delta O(\mathbf{k},\omega) = \sum_i \delta O_i(\mathbf{k},\omega) / L$ of these three local responses.

As a result of the broken translational symmetry of the central 2-by-2 unit cell supersolid, bands are back-folded. More precisely, due to the reduced translational symmetry all independent QP modes lie within the reduced first Brillouin zone ($1.\textrm{BZ}'$), as given by the reciprocal lattice vectors $\mathbf{G}_r$ defined in Section~\ref{sec:Commutator_deviation}. We thus find one gapless band with four Nambu-Goldstone cones in the full $1.\textrm{BZ}$ of the optical lattice, associated with the condensate fraction and its fluctuation. Furthermore, there are multiple gapped amplitude-dominated modes which are folded back equally and exhibit avoided crossings at degeneracy points. The amplitude modes are furthermore mixed with density fluctuations and where density fluctuations are small, the condensate fraction fluctuates at an approximately fixed particle density (see Fig.~\ref{fig:QP_response}$a$ and $c$). 

\section[D]{The roton minimum}

\begin{figure}[h]
 \centering
% \captionsetup{width=0.95\textwidth}
 \includegraphics[width=0.42\columnwidth]{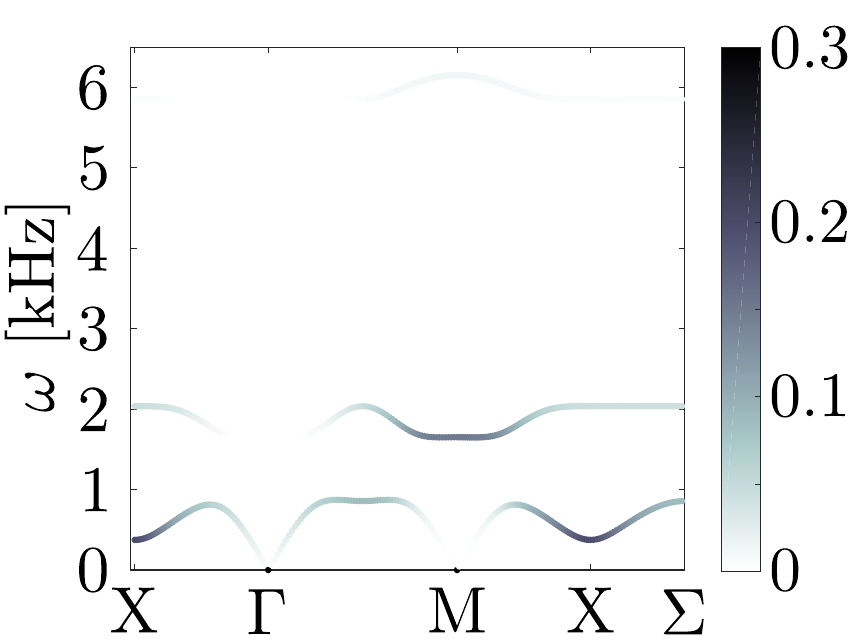}
 \caption{Dynamic structure factor along high symmetry points of a square lattice for a self-consistent checkerboard MF ground-state at fixed lattice filling $n = 0.45$ with all other paramters as in the main part.}
 \label{fig:QP_roton}
\end{figure}

In contrast to off-resonant dressing the van-der-Waals interaction potential relevant at near-resonant Rydberg excitation does not dictate a length scale by itself. Instead it is the underlying optical lattice in combination with the spontaneously broken lattice translation symmetry that determines the position of the roton minimum. Within the quasiparticle method we can visualize the roton minimum associated with the roton-instability, for example, leading to the formation of a 2-by-2 unit cell supersolid. To do so we enforce a checkerboard (CB) mean-field state via a corresponding choice of boundary conditions. While this state my not necessarily be a mean-field ground-state, it can still serve as a quasi-vacuum state of the QP theory. Here we consider a fixed lattice filling of $n = \sum_{\sigma,i} n_{\sigma i} / L = 0.45$, used in place of $\mu$ in the main part. Indeed, we find a roton minimum at the $X$ points of the $1.\textrm{BZ}$ supporting a maximum in the dynamic structure factor (see Fig.~\ref{fig:QP_roton}). Note that also for this reduced filling the ground-state energy $E_{\textrm{MF+QP}}$ of the 2-by-2 unit cell is below the corresponding value for the CB case, while there is no QP instability as all QP energies, including those of the roton, are purely real valued. Instead, the fraction of modes in the Gutzwiller excitations (small if the QP theory applies well) $\epsilon = 8.7 \%$ of the CB is more than 2.5 times $\epsilon = 3.3 \%$ of the 2-by-2 unit cell.

%\begin{thebibliography}

\bibliography{library}

%\end{thebibliography}

\end{document}